\documentclass[
reprint,
superscriptaddress,
showpacs,
preprintnumbers,
nofootinbib,
amsmath,amssymb,
aps,prd
floatfix
]{revtex4-1}
\usepackage[caption=false]{subfig}

\usepackage{subfig}
\usepackage{blindtext, subfig}
\usepackage{graphicx}
\usepackage{dcolumn}
\usepackage{bm}
\usepackage{hyperref}
\usepackage[capitalise]{cleveref}
\usepackage[mathlines]{lineno}
\usepackage{lipsum}
\usepackage{multicols}
\usepackage{xcolor}

\begin{document}

\preprint{ICPP-027}

\title{Connecting the Muon Anomalous Magnetic Moment and the Multi-lepton Anomalies at the LHC}

\author{Danielle Sabatta}
\email{danielle.dorinda.sabatta@cern.ch}
\affiliation{School of Physics and Institute for Collider Particle Physics,
University of the Witwatersrand, Johannesburg,
Wits 2050, South Africa.}
\author{Alan S. Cornell}
\email{acornell@uj.ac.za }
\affiliation{Department of Physics, University of Johannesburg, PO Box 524, Auckland Park 2006, South Africa.}
\author{Ashok Goyal}
\email{agoyal45@yahoo.com}
\affiliation{School of Physics and Institute for Collider Particle Physics,
University of the Witwatersrand, Johannesburg,
Wits 2050, South Africa.}
\author{Mukesh Kumar}
\email{mukesh.kumar@cern.ch}
\affiliation{School of Physics and Institute for Collider Particle Physics,
University of the Witwatersrand, Johannesburg,
Wits 2050, South Africa.}
\author{Bruce Mellado}
\email{bmellado@mail.cern.ch}
\affiliation{School of Physics and Institute for Collider Particle Physics,
University of the Witwatersrand, Johannesburg,
Wits 2050, South Africa.}
\affiliation{iThemba LABS, National Research Foundation, PO Box 722, Somerset West 7129, South Africa.}
\author{Xifeng Ruan}
\email{xifeng.ruan@cern.ch}
\affiliation{School of Physics and Institute for Collider Particle Physics,
University of the Witwatersrand, Johannesburg,
Wits 2050, South Africa.}

\begin{abstract}
{A number of predictions were made in Ref.~\cite{vonBuddenbrock:2016rmr} pertaining to the anomalous production of multiple leptons at the Large Hadron Collider (LHC).  Discrepancies in multi-lepton final states have now become statistically compelling with the available Run~2 data. These could be connected with a heavy boson, $H$,  decaying predominantly into a SM Higgs boson, $h$, and a singlet scalar, $S$, where $m_H\approx 270$\,GeV and $m_S\approx 150$\,GeV. These can be embedded into a scenario where a Two Higgs Doublet is considered with an additional singlet scalar, 2HDM+S. The long-standing discrepancy in the muon anomalous magnetic moment, $\Delta a_\mu$, is interpreted in the context of the 2HDM+S type-II and type-X, along with additional fermionic degrees of freedom. The 2HDM+S model alone with the constraints from the LHC data does not seem to explain the $\Delta a_\mu$ anomaly. However, adding  fermions with mass of order $\mathcal{O}(100)$\,GeV can explain the discrepancy for low enough values of fermion-scalar couplings.} 
\end{abstract}
\maketitle

\section{\label{sec:intro}Introduction}

The anomalous magnetic moment of the muon, $a_\mu = (g-2)_\mu/2$, is one of the most important precision observables to test the Standard Model (SM) and provide a complementary, non-collider constraint of beyond the Standard Model (BSM) physics. Currently, the discrepancy between the experimental measurement and the SM prediction is $\sim 3.5 \sigma$~\cite{Bennett:2006fi, Bennett:2004pv, Bennett:2002jb, Brown:2001mga, Jegerlehner:2009ry, Hagiwara:2011af, Davier:2010nc, Blum:2013xva}, where:
\begin{equation}\label{eqn:discrepancy}
\Delta a_\mu = a_\mu^{\rm Exp} - a_\mu^{\rm SM} = 2.87 (80) \times 10^{-9}.
\end{equation} 
This opens a window of opportunity for quantum corrections driven by BSM particles~\cite{Lindner:2016bgg, Freitas:2014pua, Queiroz:2014zfa, Biggio:2014ela, Biggio:2016wyy}.  In a model independent scenario a detailed study~\cite{Freitas:2014pua} shows the contribution to $a_\mu$ for the BSM particles of masses of a few 100~GeV. A complete two-loop contribution to $a_\mu$ in the two-Higgs doublet model (2HDM) is performed in Refs.~\cite{Broggio:2014mna, Cherchiglia:2016eui} to explain the anomaly, $\Delta a_\mu$, which constrains the parameter space of the model. These studies connect the $\Delta a_\mu$ with the collider studies at the Large Hadron Collider (LHC) and the Fermilab experiments.

In this paper we shall draw upon our previous works, where we had studied the effects of a new scalar, $H$, heavier than the SM Higgs, as related to Run~1 results from the LHC~\cite{vonBuddenbrock:2015ema,vonBuddenbrock:2016rmr}. From an effective Lagrangian approach, the best fit mass of $H$ was determined as $m_H=272^{+12}_{-9}$\,GeV, where these previous works drew on (but were not limited to) the production of multiple leptons (in association with $b$-quarks) as had been studied in the searches for the SM Higgs. Note that these studies had been in association with the top quark. As a result of our earlier studies, it became necessary to introduce a scalar mediator, $S$, such that our effective vertices were constructed of $HSh$, $HSS$, and $Hhh$ interactions. Furthermore, the $S$ could decay (in a Higgs-like manner) to SM particles~\cite{vonBuddenbrock:2016rmr}.

Furthermore, we made a number of predictions related, at high energy proton-proton collisions, to the production of multiple leptons in Refs.~\cite{vonBuddenbrock:2015ema,vonBuddenbrock:2016rmr}.
Assuming that the singlet scalar behaves like a SM Higgs-like boson, the data can be described with $m_H\approx 270$\,GeV and $m_S\approx 150$\,GeV. These discrepancies have now become statistically compelling with the available Run~2 data~\cite{vonBuddenbrock:2019ajh} where the mass points and parameters were fixed from our earlier studies, and as such were not altered in our model to better explain the data.
The final states were selected as per the predictions in Refs.~\cite{vonBuddenbrock:2015ema,vonBuddenbrock:2016rmr}, which predate the Run~2 data. These include the anomalous production of opposite-sign, same-sign and three leptons  in the presence and absence of  $b$-quarks. 

The discrepancies which arise in final states and regions of the phase space where different processes dominate in the SM description do not point to a likely mis-modeling of a specific SM process. Rather the anomalies
and their kinematic characteristics are reasonably well described by a simple ansatz, where $H\rightarrow Sh$ is produced via gluon-gluon fusion and in association with top quarks in high-energy proton-proton collisions. It is, therefore, appropriate to understand the possible connection of the above mentioned spectroscopy with the $\Delta a_\mu$ anomaly through loop corrections. 

The above mentioned $H$ and $S$ can be embedded into a 2HDM scenario with an additional scalar, where $S$ is a singlet under the SM gauge groups~\cite{vonBuddenbrock:2016rmr,Muhlleitner:2016mzt,vonBuddenbrock:2018xar}. This was done explicitly in Ref.~\cite{vonBuddenbrock:2018xar}, where a study of this embedding's 2HDM+S parameter space was made
that can accommodate the discrepancies between the SM and the data reported in Ref.~\cite{vonBuddenbrock:2017gvy}. Here we investigate whether the 2HDM+S model with the parameter space identified in Ref.~\cite{vonBuddenbrock:2017gvy} can account for the $\Delta a_\mu$ anomaly or whether new degrees of freedom are necessary.

While the multi-lepton anomalies reported in Refs.~\cite{vonBuddenbrock:2017gvy,vonBuddenbrock:2019ajh} seem to be fairly well described with the simple ansatz mentioned above, in Ref.~\cite{vonBuddenbrock:2019ajh} a more complex picture was indicated in the data than this 2HDM+S model. The specific processes which indicated this greater complexity were in the dilepton system, namely the invariant mass, the transverse mass and the missing transverse energy. These tended to be wider than what is predicted
by the $S\rightarrow W^+W^-\rightarrow \ell^+\ell^- ~(\ell=e,\mu)$ decay chain. New leptonic degrees of freedom could significantly alter the decays of $S$, thus modifying the differential distribution predicted by the model~\cite{vonBuddenbrock:2017gvy}. In this light, we explore what one can learn from the $\Delta a_\mu$ anomaly with regards to these potential new degrees of freedom.

In this short article we connect $\Delta a_\mu$ with the constrained parameter space of the 2HDM+S at the LHC. We briefly explain the model considered for this study in Section~\ref{sec:2hdm+s}, along with the constraints on the parameter space from previous studies. The one- and two-loop formulae are discussed in Section~\ref{sec:2hdm+s_cons}, and results of the study are detailed in Section~\ref{sec:result}. In Section~\ref{constraint} the implications of this study to other processes is also discussed, where finally, a summary and conclusion of this study is presented in Section~\ref{sec:concl}.

\section{The Model}
\label{sec:2hdm+s}
As mentioned in Section~\ref{sec:intro}, we are considering the 2HDM+S model as a possible explanation for $\Delta a_\mu$. Following Refs.~\cite{vonBuddenbrock:2018xar, Muhlleitner:2016mzt, Ivanov:2017dad} this model is, in brief, based on the well-known 2HDM with the addition of a real singlet scalar $S$. The potential is given by:
\begin{align}
&V(\Phi_1, \Phi_2, \Phi_S) \notag\\
&= m_{11}^2 \left|\Phi_1\right|^2 + m_{22}^2 \left|\Phi_2\right|^2 - m_{12}^2 \left(\Phi_1^\dagger \Phi_2 + {\rm h.c.}\right) \notag\\
&\, + \frac{\lambda_1}{2} \left(\Phi_1^\dagger \Phi_1\right)^2 + \frac{\lambda_2}{2} \left(\Phi_2^\dagger \Phi_2\right)^2
 + \lambda_3 \left(\Phi_1^\dagger \Phi_1\right) \left(\Phi_2^\dagger \Phi_2\right)  \notag \\
&\,+ \lambda_4 \left(\Phi_1^\dagger \Phi_2\right) \left(\Phi_2^\dagger \Phi_1\right) + \frac{\lambda_5}{2}\left[\left(\Phi_1^\dagger \Phi_2 \right)^2 + {\rm h.c.}\right] \notag \\
&\, + \frac{1}{2} m_S^2 \Phi_S^2 + \frac{\lambda_6}{8} \Phi_S^4 + \frac{\lambda_7}{2} \left(\Phi_1^\dagger \Phi_1 \right)\Phi_S^2
+ \frac{\lambda_8}{2} \left(\Phi_2^\dagger \Phi_2 \right)\Phi_S^2, \label{pot}
\end{align}
where the fields $\Phi_1$ and $\Phi_2$ are the $SU(2)_L$ Higgs doublets, while $\Phi_S$ is the singlet scalar field. The first three lines correspond to the terms in the real 2HDM potential. The final four terms relate to the singlet $S$ field. Recall that models which have more than one Higgs doublet can have tree-level Flavor Changing Neutral Currents (FCNCs). In order to avoid these tree-level currents, the usual approach is to couple all quarks with the same charge to a single double.

Due to the addition of a singlet scalar this model has three physical CP-even scalars $h, S,$ and $H$, with one CP-odd scalar $A$ and charged scalar $H^\pm$. Other parameters of this model are the mixing angles $\alpha_1, \alpha_2, \alpha_3$ and $\tan\beta$, vacuum expectation values ($vev$) $v, v_S$, and the masses $m_h, m_S, m_H, m_A, m_{H^\pm}$. As discussed in Section~\ref{sec:intro}, the masses of many of these parameters are fixed {\it a priori} from previous studies \cite{vonBuddenbrock:2016rmr,Muhlleitner:2016mzt,vonBuddenbrock:2018xar}, where the as yet constrained mass $m_A$, and to a lesser extent $m_S$, will be scanned over in this study.

The relevant Yukawa couplings between the SM fermions and 2HDM+S scalar mass eigenstates are given as:
\begin{align}
-& {\cal L}_Y^{\rm 2HDM+S} \notag \\
=&\, \sum_{f = u, d, \ell} \frac{m_f}{v} \Big[ y_f^h h \bar{f} f + y_f^H H \bar{f} f + y_f^S S \bar{f} f - i  y_f^A A \bar{f}\gamma_5 f \Big] \notag \\ 
&\qquad+ \Big[ \sqrt{2} V_{ud} H^+ \bar{u} \left(\frac{m_u}{v} y_u^A P_L + \frac{m_d}{v} y_d^A P_R\right) \notag\\
 &\qquad \qquad\qquad\qquad+\sqrt{2} \frac{m_\ell}{v} y_\ell^A H^+ \bar{\nu} P_R \ell + {\rm h.c.}\Big].
\label{lag}
\end{align}
For the details on the couplings and other information we refer the readers to Refs.~\cite{vonBuddenbrock:2018xar, Muhlleitner:2016mzt}. Furthermore, for our studies we only consider type-II and lepton-specific (type-X) models within the parameter space considered in Ref.~\cite{vonBuddenbrock:2018xar}. 

\begin{figure}[t]
   \centering
   \subfloat[]{\includegraphics[width=0.5\linewidth]{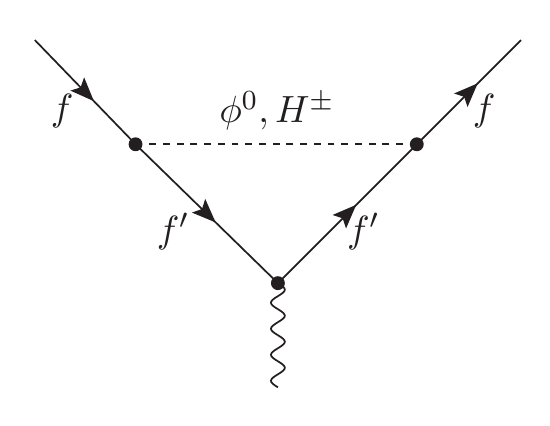}{\label{1loop}}}
   \subfloat[]{\includegraphics[width=0.5\linewidth]{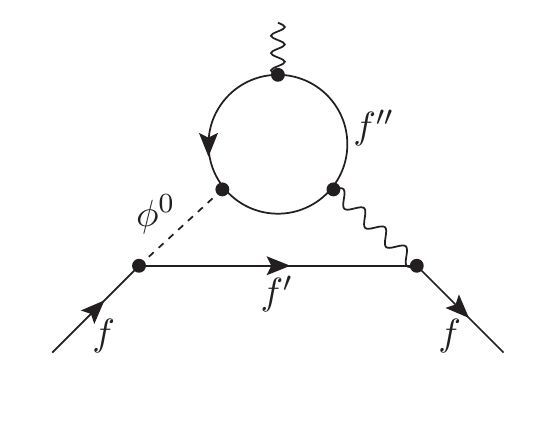}{\label{2loop}}}
 \caption{Representative (a) one-loop and (b) two-loop diagrams contributing to the $\Delta a_\mu$. For 2HDMs, $\phi^0 = h, H, A$ while in the case of the 2HDM+S, $\phi^0$ also gets a contribution from $S$. In a 2HDM or 2HDM+S scenario, the fermions $f$ and $f^\prime$ can be considered as the SM leptons, however $f^{\prime\prime}$ could be quarks and leptons. The dominant contributions comes from $ f^{\prime\prime} = t, b, \tau$. For 2HDM+S+f model, $f^\prime$ could be taken as BSM charged fermions with neutral scalars.}
    \label{1&2-loop}
\end{figure}
\begin{figure*}[t]
\centering
\subfloat[]{\includegraphics[width=.49\linewidth]{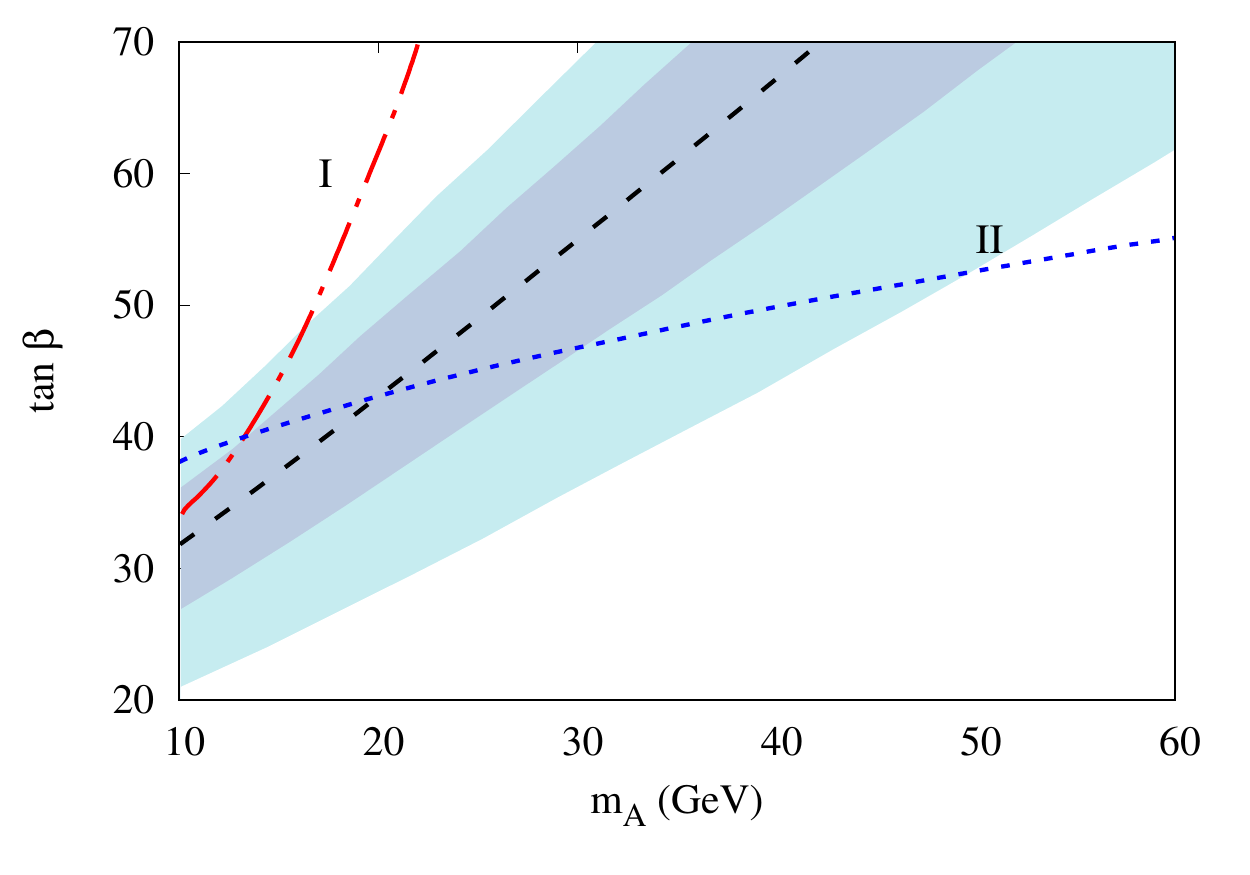}\label{2HDM_Type_II}} 
\subfloat[]{\includegraphics[width=.49\linewidth]{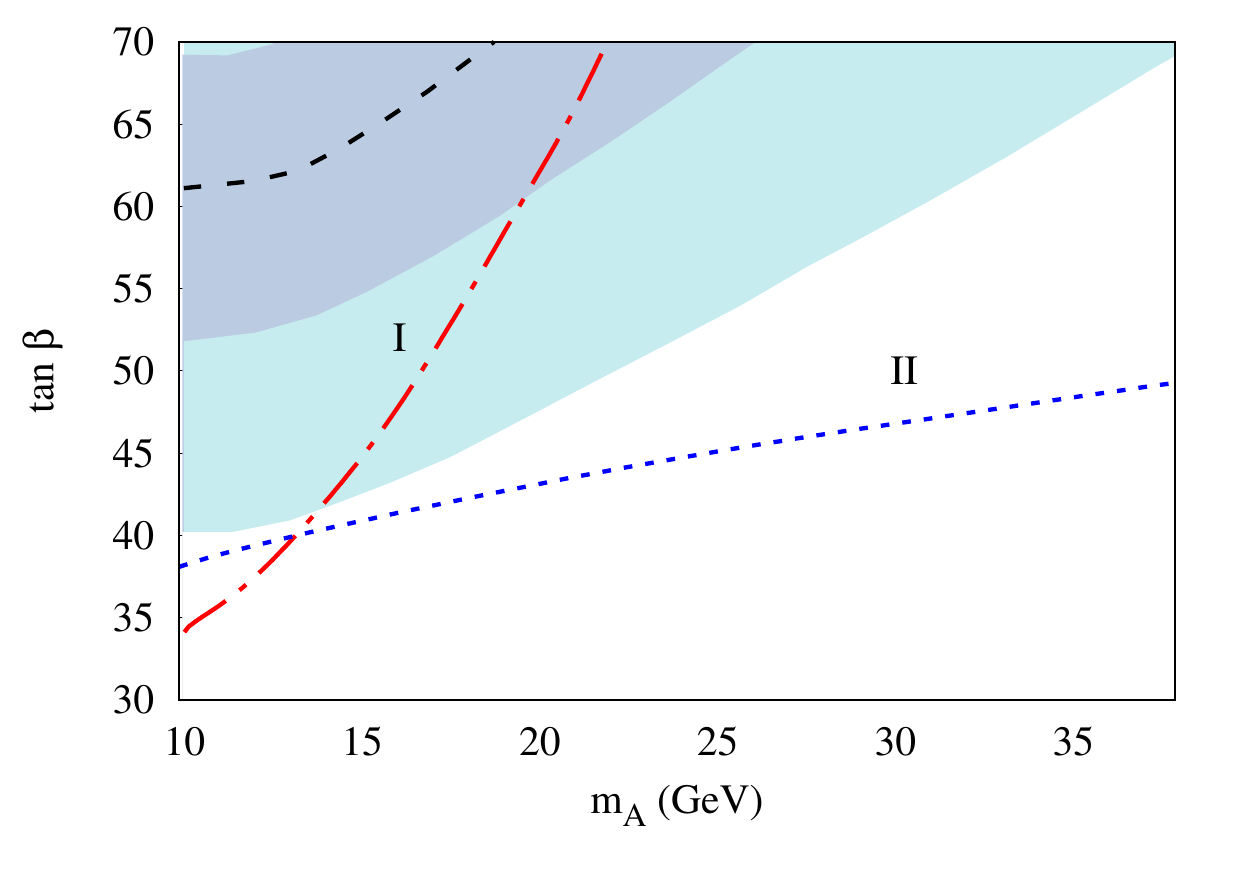}\label{2HDM_Type_X}}
\caption{Values of $\tan\beta$ and $m_A$ required to explain the $\Delta a_\mu$ in (a) type-II and (b) type-X 2HDM. The one and two sigma regions are shown with dark blue and light blue, respectively. This parameter space can be further constrained by experimental data. The regions I and II (above the dashed {\it red} and {\it blue} lines respectively) are excluded by the direct search at the LEP and the $\tau$ decay ($\tau \to \mu \nu_\mu \nu_\tau$) respectively at 95\% C.L~\cite{Abe:2015oca}.}
\label{2HDM}
\end{figure*}
\begin{figure*}[t]
\centering
 \subfloat[]{\label{2HDM+S_Type_II}\includegraphics[width=.49\linewidth]{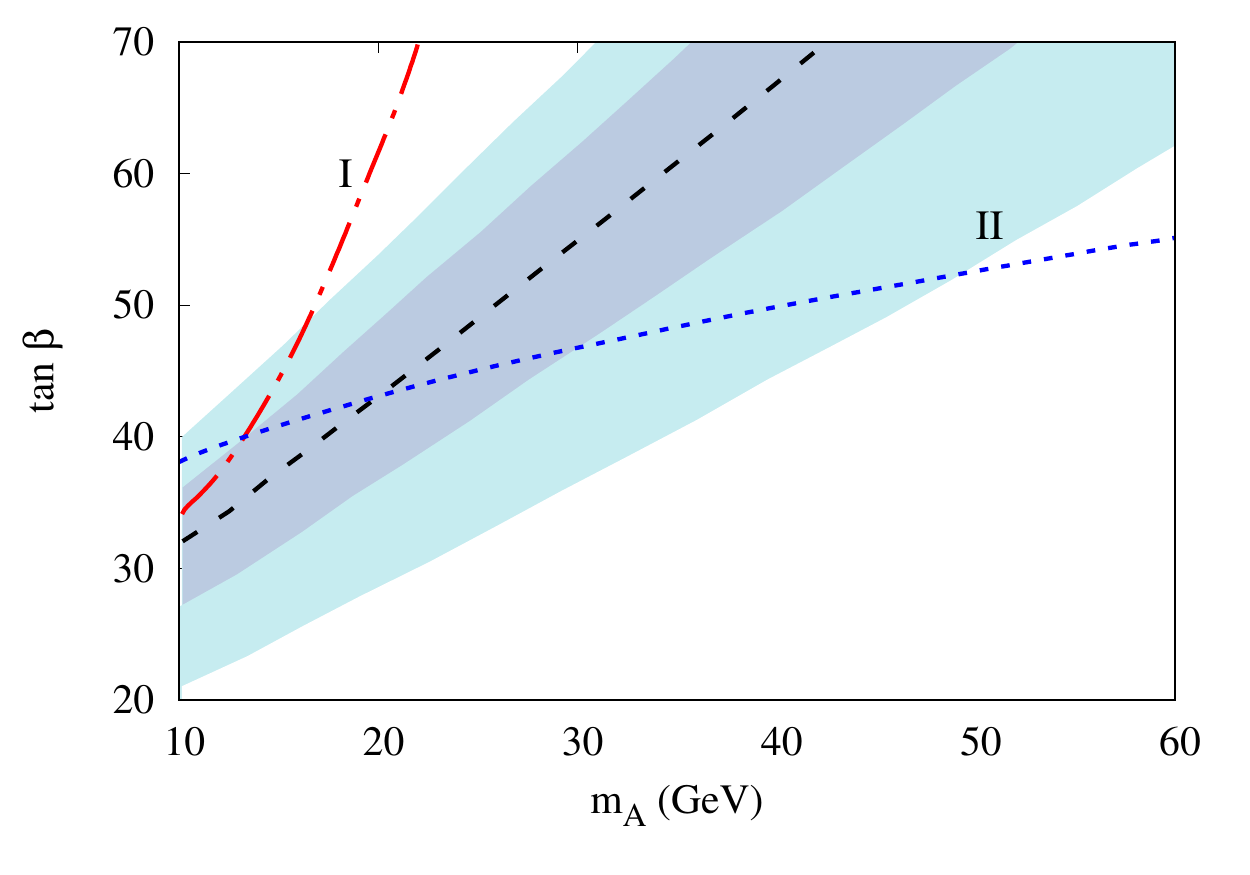}} 
  \subfloat[]{\label{2HDM+S_Type_X}\includegraphics[width=.49\linewidth]{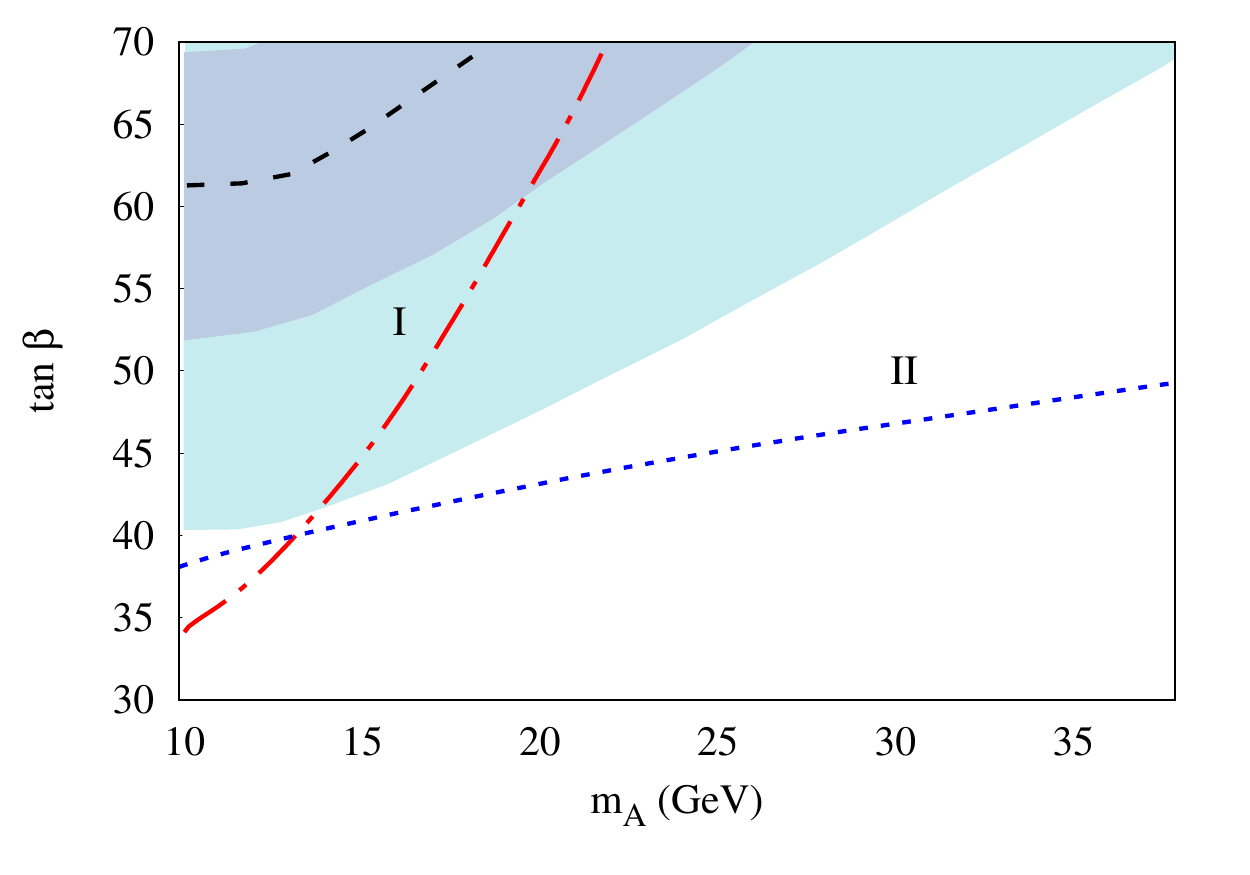}}
   \caption{Values of $\tan\beta$ and $m_A$ required to explain the $\Delta a_\mu$ in (a) type-II and (b) type-X 2HDM+S. The one and two sigma regions are shown with dark blue and light blue, respectively. Note the similarity to the parameter space required by the 2HDM without $S$ in~\cref{2HDM}. The regions I and II (above the dashed {\it red} and {\it blue} lines respectively) are excluded by the direct search at the LEP and the $\tau$ decay ($\tau \to \mu \nu_\mu \nu_\tau$) respectively at 95\% C.L~\cite{Abe:2015oca}.}
  \label{2HDM+S_high_tanbeta}
\end{figure*}
  
Since these types of models are highly constrained by many studies, we also looked at a scenario by adding BSM leptons, which are singly charged. Specifically, we consider light leptons with a mass of ${\cal O}(10^2)$~GeV that are not produced {\it directly} at colliders. This means that these leptons are to be treated as mediators, and would contribute via loop corrections to the $\Delta a_\mu$ anomaly. For the sake of simplicity, and without any loss of generality, we consider singly charged SM singlet vector-like leptonic fermions with chiral components which transform as follows~\cite{Freitas:2014pua}:
\begin{align}
{\cal L} \supset -y_{f^\prime}^S {\overline {l_R}} \Phi_S f_L^{\prime} - \sum_{i=1}^2 y_{f^\prime}^i {\overline {L_l}} \Phi_i f_R^{\prime}  + {\rm h.c.}, \label{lagsf}
\end{align}     
where $l_{R}$ and $L_l$ are the SM singlet and doublet leptons, and $f^{\prime}_{L/R}$ are the BSM singly charged vector-like leptons with left and right chirality. That is, under SM gauge transformations the different chiral components transform the same way. The interaction Lagrangian,~\cref{lagsf}, can now be easily cast in terms of the scalar mass eigenstates as in~\cref{lag}. In adding these interactions to the 2HDM+S we have expanded our model to what we shall label as a 2HDM+S+f model. However, these fermions are also constrained by collider searches in terms of masses and model-dependent couplings, which we further explain in section~\ref{constraint}. 

Note that the overall coupling should be constrained as $y_{f, f^\prime}^i\le \sqrt{4\pi}$, though it should be noted that all the couplings which appear in the interactions are the functions of the mixing angles $\alpha_i$ and $\beta$ used to diagonalize the mass matrix appropriately in the model. Without loss of generality, we can take $-\pi/2 \leq \alpha_i \leq \pi/2$ and we scan over $\beta$ in the coming sections, along with the mass of the new vector-like fermion, $f$.   

\begin{figure*}[t]
\centering
\subfloat[] {\includegraphics[width=.49\linewidth]{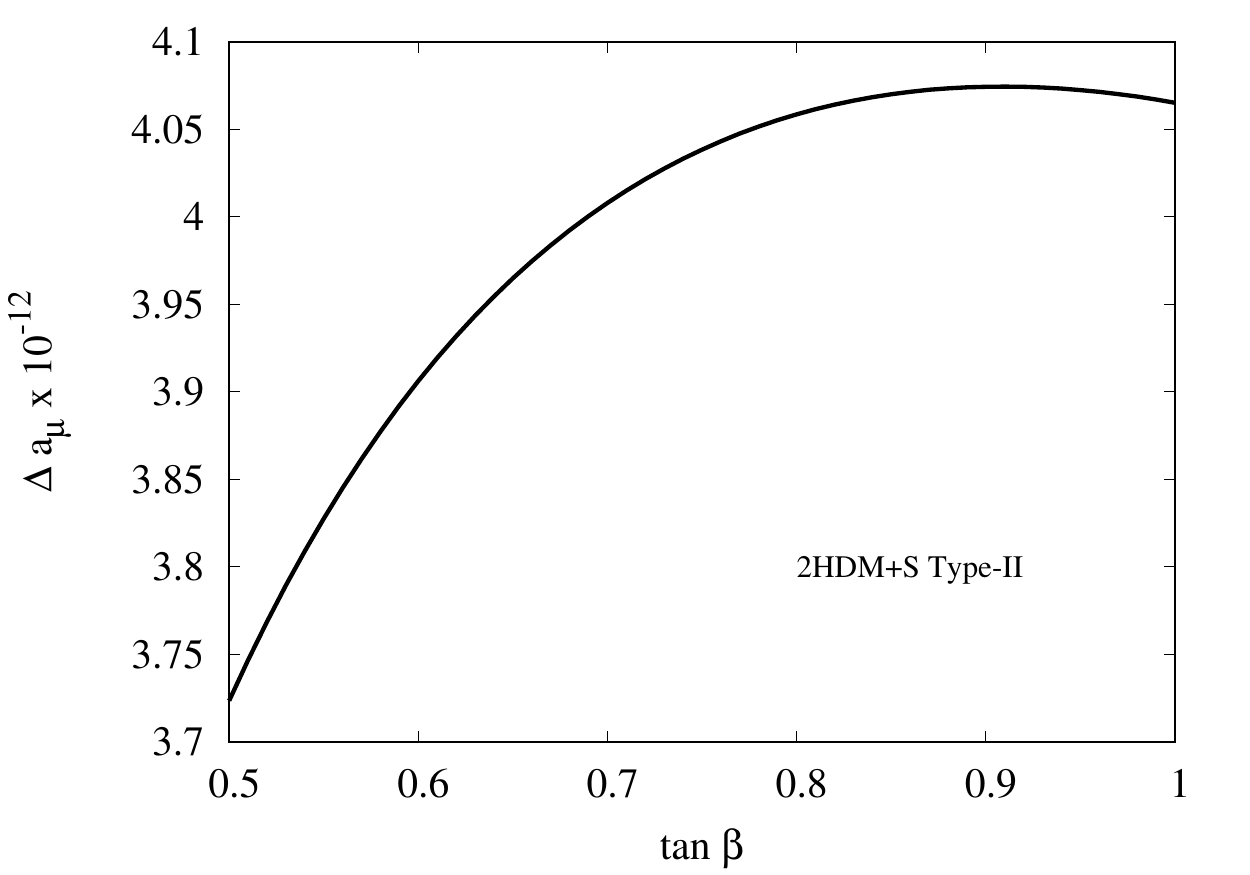}\label{2HDM+S_Type_II_contribution}}
\subfloat[]{\includegraphics[width=.49\linewidth]{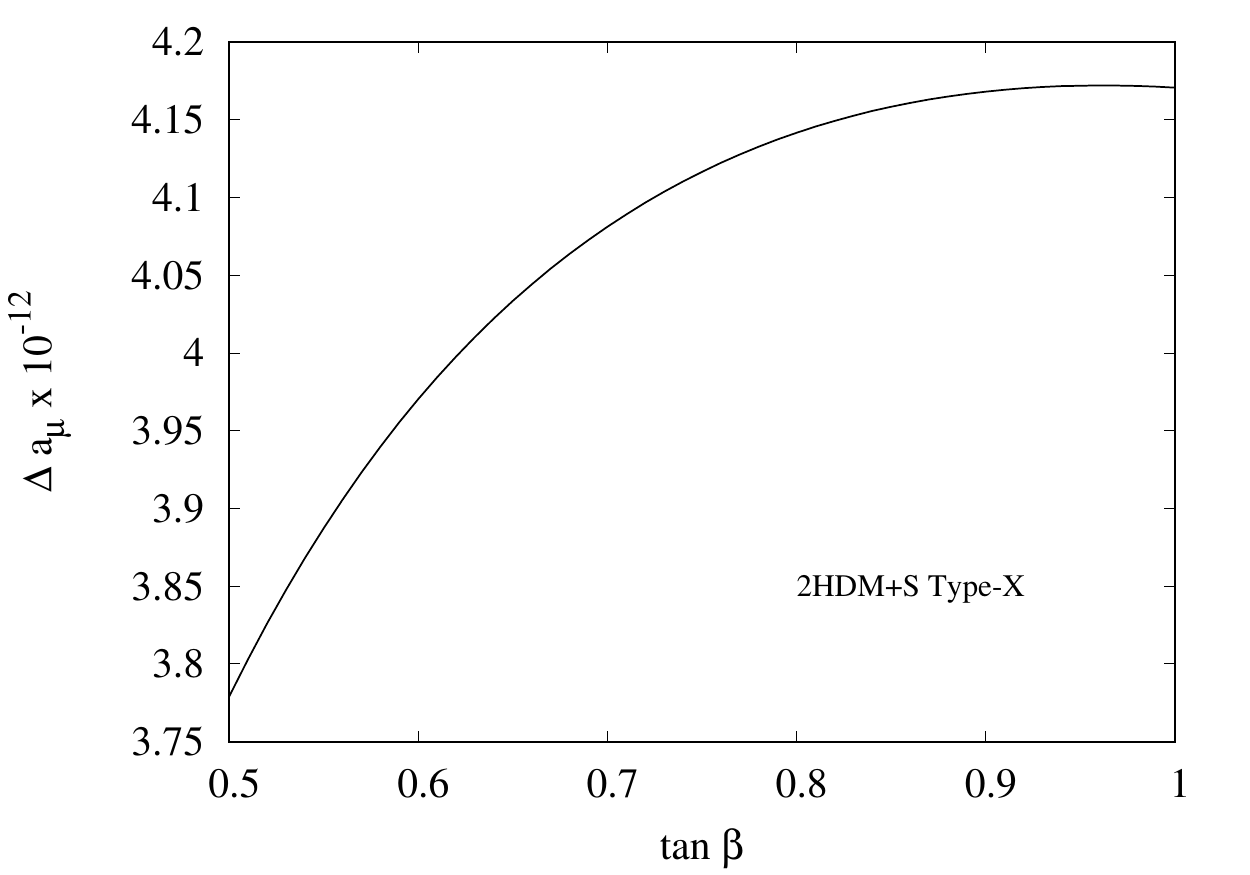}\label{2HDM+S_Type_X_contribution}}
\caption{One- and two-loop contributions from the 2HDM+S to $\Delta a_\mu$ using~\cref{eqn:1-loop_SM+S} and~\cref{eqn:2-loop_SM+S} in (a) type-II and (b) type-X 2HDM+S. Here $m_A = 600$~GeV is taken.}
   \label{2hdm+s_cont}
\end{figure*}

\begin{figure*}[t]
\centering
   \subfloat[]{\includegraphics[width=.49\linewidth]{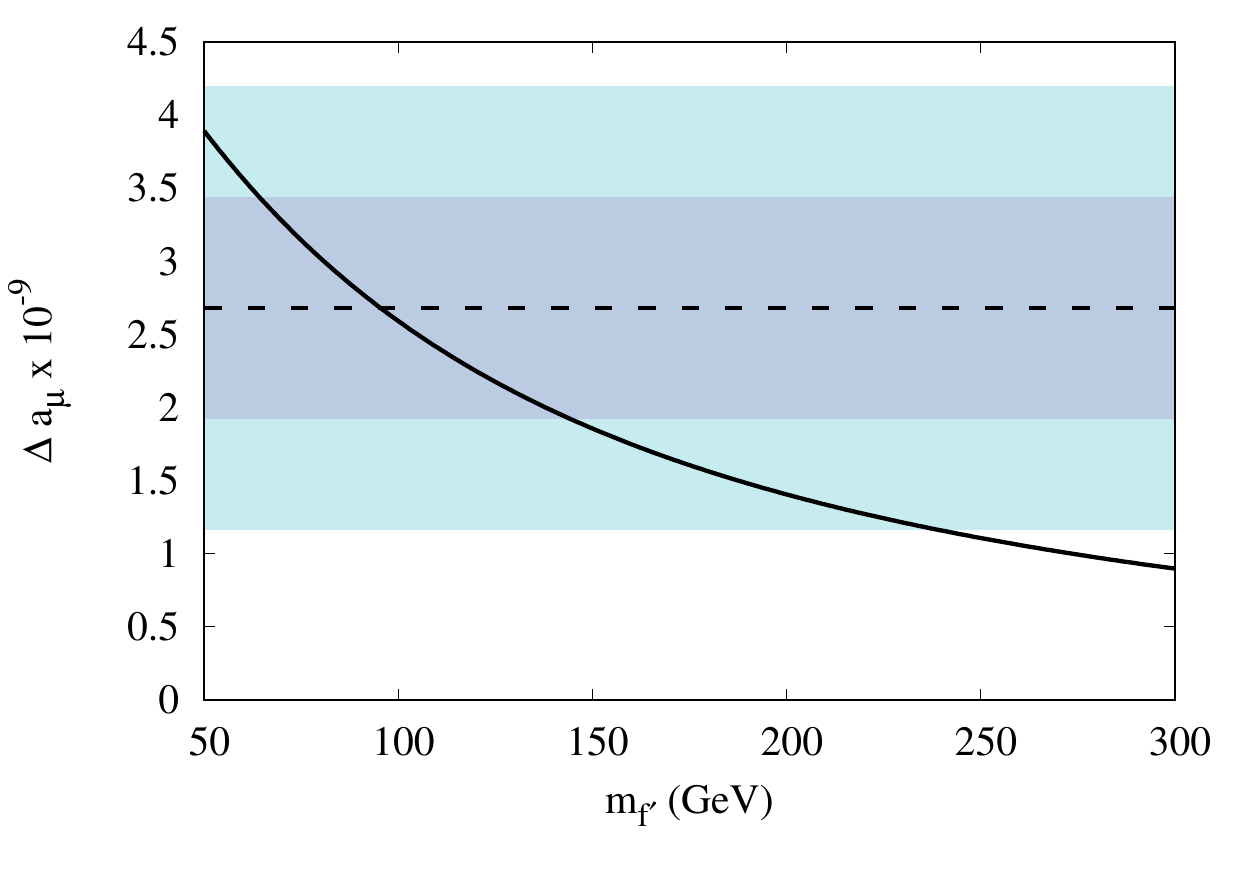}\label{100f_cont}} 
   \subfloat[]{\includegraphics[width=.49\linewidth]{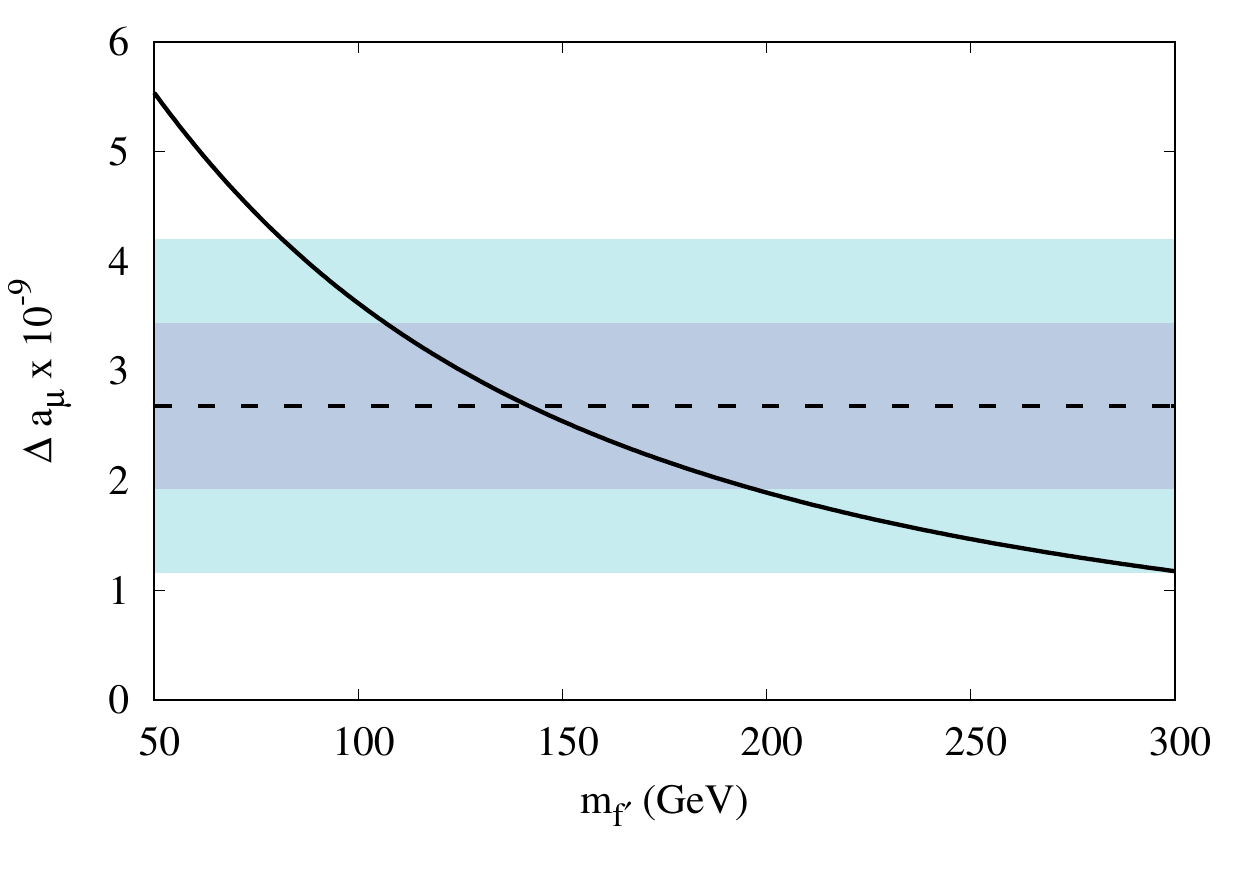}\label{150f_cont}}\\
   \subfloat[]{\includegraphics[width=.49\linewidth]{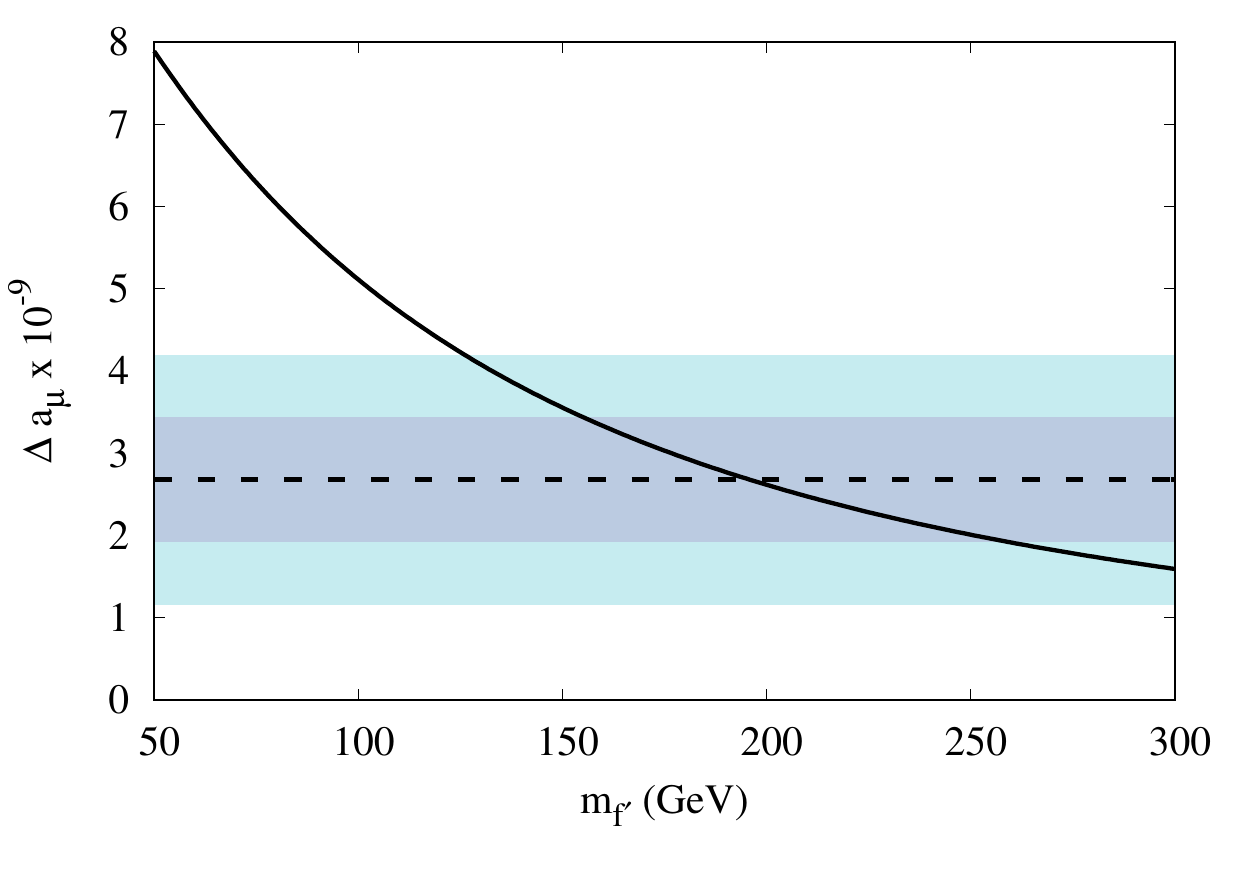}\label{200f_cont}}
   \subfloat[]{\includegraphics[width=.49\linewidth]{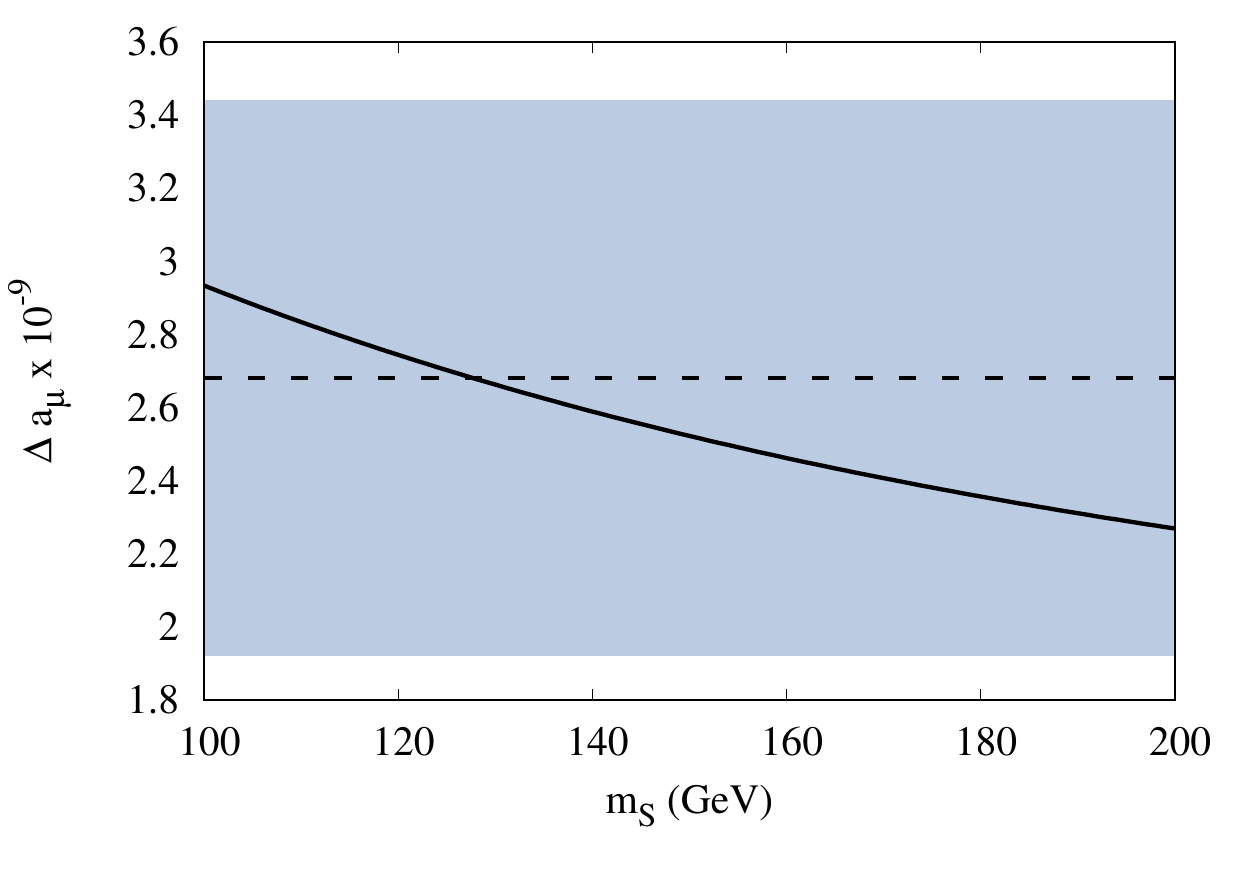}\label{mSvary}}  
   \caption{Contributions to $\Delta a_\mu$ from the one-loop diagram in~\cref{1&2-loop} by (a), (b), (c) varying fermion masses and (d) varying mass of $S$. The dashed line shows the value of the discrepancy as given in~\cref{eqn:discrepancy}, and the one and two sigma regions are shown with dark blue and light blue, respectively. The fermion couplings are fixed at (a) $y^h_{f^\prime}=1$, $y^A_{f^\prime}=3.5$, $y^S_{f^\prime}=y^H_{f^\prime}=1.5$, (b) $y^h_{f^\prime}=1$, $y^A_{f^\prime}=3.5$, $y^S_{f^\prime}=y^H_{f^\prime}=2$, (c) $y^h_{f^\prime}=1$, $y^A_{f^\prime}=3.5$, $y^S_{f^\prime}=y^H_{f^\prime}=2.5$, (d) $y^h_{f^\prime}=1$, $y^A_{f^\prime}=3.5$, $y^S_{f^\prime}=y^H_{f^\prime}=1.5$, and $m_{f^\prime}=100$~GeV.}
   \label{fermion_cont}
\end{figure*}

\begin{figure*}[t]
\centering
   \subfloat[]{\includegraphics[width=.49\linewidth]{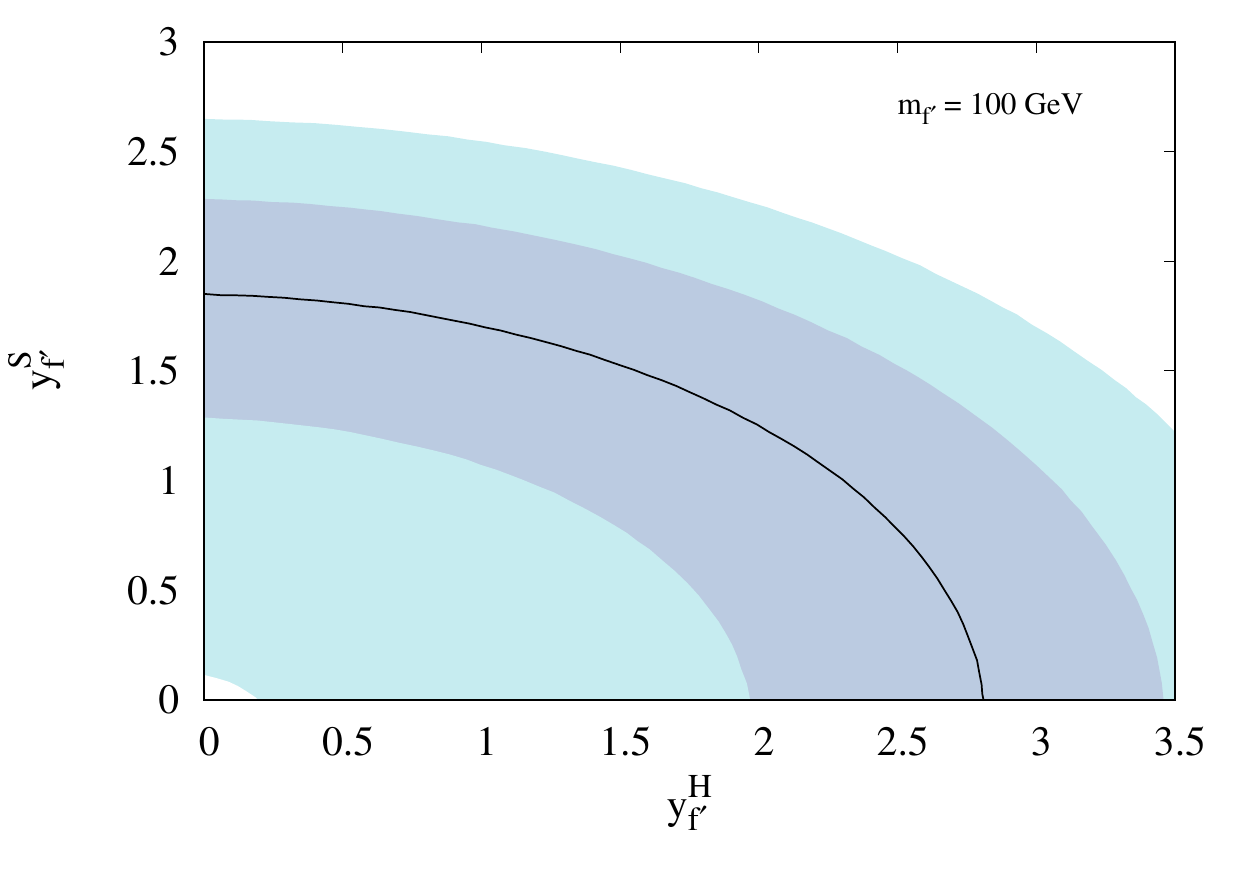}\label{100f_couplings}}
   \subfloat[]{\includegraphics[width=.49\linewidth]{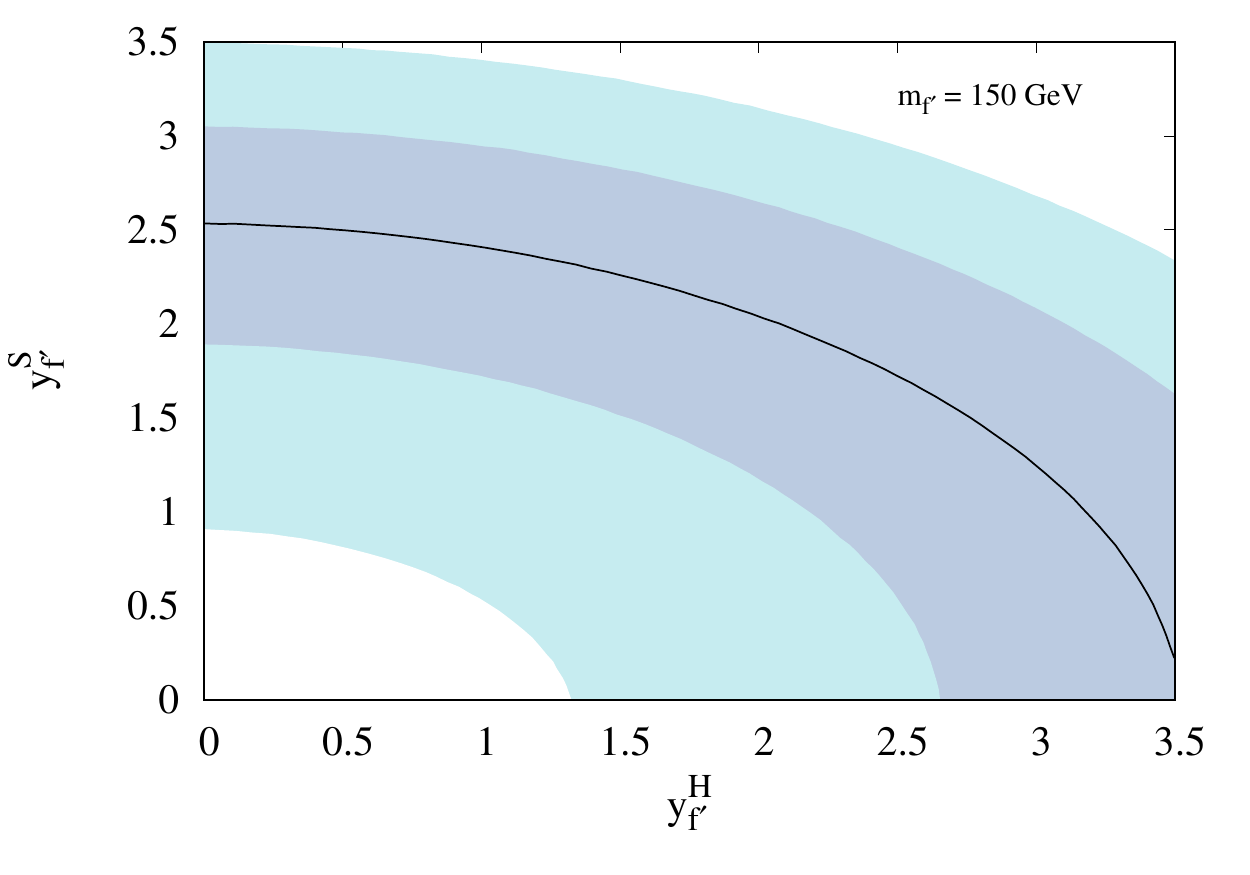}\label{150f_couplings}}\\
   \subfloat[]{\includegraphics[width=.49\linewidth]{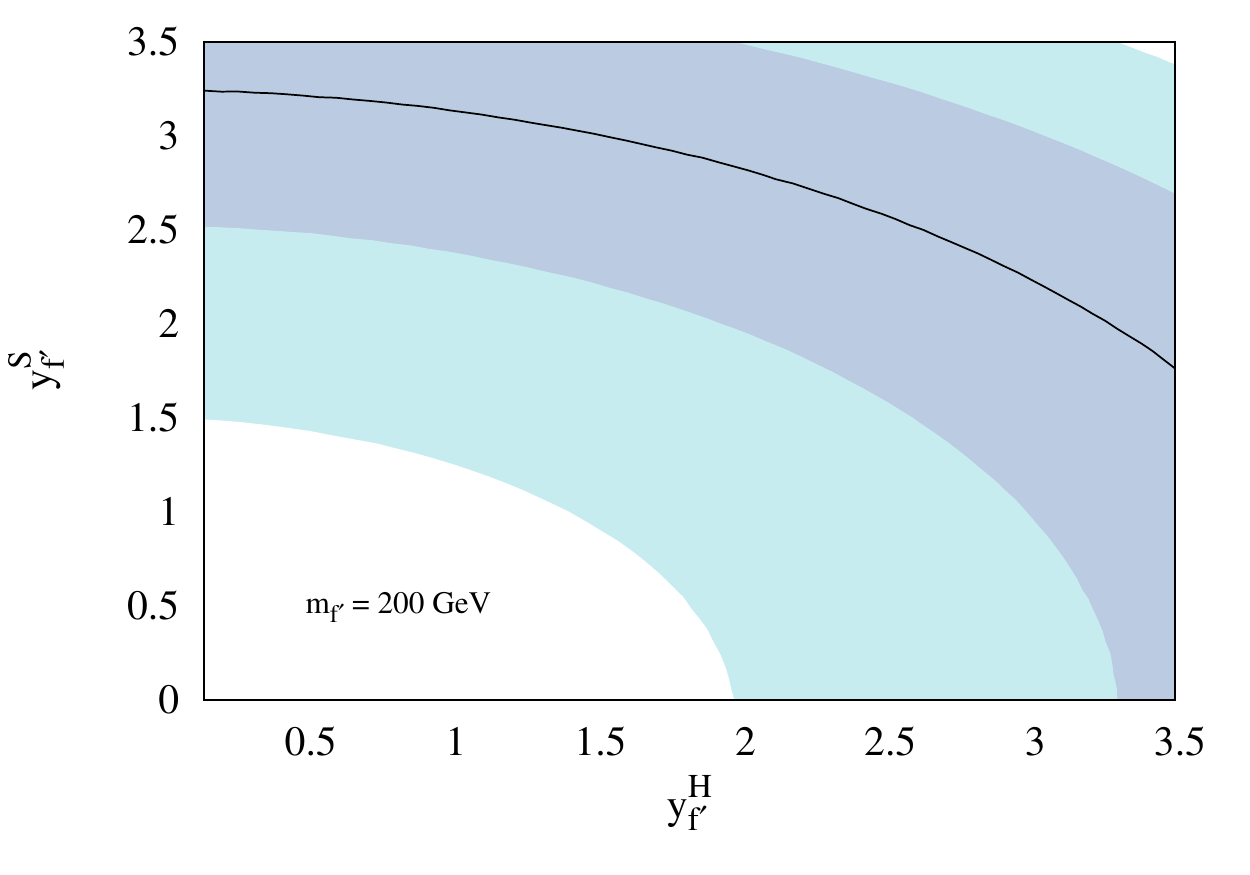}\label{200f_couplings}}
   \caption{Some of the fermion coupling values required for different masses $m_{f^\prime}$. The masses are kept constant at (a) $m_{f^\prime}=100$~GeV, (b) $m_{f^\prime}=150$~GeV, and (c) $m_{f^\prime}=200$~GeV.}
   \label{fermion_couplings}
\end{figure*}
 
\section{Contributions to $\Delta a_\mu$}
\label{sec:2hdm+s_cons}
The 2HDM contributions to the $\Delta a_\mu$ have been calculated and are known up to the two-loop level~\cite{Broggio:2014mna, Cherchiglia:2016eui}, where these calculations also apply for the 2HDM+S with appropriate coupling arrangements. The one- and two-loop diagrams contributing to $\Delta a_\mu$  are shown in ~\cref{1&2-loop}. It has been shown that the type-II and type-X (lepton specific) 2HDMs are suitable to explain the discrepancy with positive contributions to the $\Delta a_\mu$. In these models, the lepton couplings to the new bosons are enlarged, while the top Yukawa coupling are kept favorably small. 

The one loop contribution from the neutral and charged scalars is given by the expression: 
\begin{equation}\label{eqn:1-loop_SM+S}
    \Delta a_\mu(1\; \textrm{loop}) = \frac{G_F m^2_\mu}{4\pi^2 \sqrt{2}} \: \sum_{j}(y^j_\mu)^2 \: r^j_\mu \: f_j (r^j_\mu),
\end{equation}
where $j=\{h, S, H, A, H^\pm \}$, $r^j_\mu = m^2_\mu / M^2_j$, and
\begin{equation}\label{eqn:f-h,S,H}
    f_{h,S,H}(r)=\int^{1}_{0} dx\frac{x^2(2-x)}{1-x+rx^2},
\end{equation}
\begin{equation}\label{eqn:f-A}
     f_{A}(r)=\int^{1}_{0} dx\frac{-x^3}{1-x+rx^2},
\end{equation}
\begin{equation}\label{eqn:f-H+-}
     f_{H^\pm}(r)=\int^{1}_{0} dx\frac{-x(1-x)}{1-(1-x)r}.
\end{equation}
The two loop contribution from the neutral scalars are given by the expression:
\begin{equation}\label{eqn:2-loop_SM+S}
    \Delta a_\mu(2\; \textrm{loop}) = \frac{G_F m^2_\mu}{4\pi^2 \sqrt{2}}\: \frac{\alpha_{em}}{\pi} \: \sum_{i,f} N^c_f Q^2_f y^i_\mu y^i_f \: r^i_f \: g_i (r^i_f),
\end{equation}
where $i=\{h, S, H, A\}$, $f=\{t, b, \tau\}$, $r^i_f=m^2_f / M^2_i$, and $m_f$, $Q_f$, and $N^c_f$ are the mass, charge, and number of color degrees of freedom of the fermion in the loop. The functions $g_i(r)$ are:
\begin{equation}\label{eqn:gi}
    g_i(r)= \int^{1}_{0} dx \frac{\mathcal{N}_i (x)}{x(1-x)-r}\ln{\frac{x(1-x)}{r}},
\end{equation}
where $\mathcal{N}_{h,S,H} (x)=2x(1-x)-1$ and $\mathcal{N}_A (x)=1$.

In this study we shall go one step further, and use the 2HDM+S+f model discussed in Section~\ref{sec:2hdm+s}, where the addition of BSM fermions gives a one-loop contribution to $\Delta a_\mu$ as given by~\cite{Freitas:2014pua}: 
\begin{equation}
\Delta a_{\mu}^{f^\prime} ({\rm 1~loop}) = \frac{1}{16\pi^2}\: \sum_{i}(y^i_{f^\prime})^2 \: r^i_\mu \: F_i (r^i_{f^\prime}),
\label{eqn:delta_a_f}
\end{equation}
where $i=\{h, S, H, A\}$, $r^i_{f^\prime}=m^2_{f^\prime} / M^2_i$, and $r^i_\mu=m^2_\mu / M^2_i$. The function $F_i$ is defined as:
\begin{equation}\label{loop_function_F}
F_i(r) = \frac{r^{3}-6r^{2}+3r+6r\ln(r)+2}{6(1-r)^{4}}.
\end{equation}
We will now use these formulae, inputting the numerical values of parameters from previous studies \cite{vonBuddenbrock:2016rmr,Muhlleitner:2016mzt,vonBuddenbrock:2018xar}, to generate the $\Delta a_\mu$, scanning across the parameters $m_f$, $m_A$, and the mixing angles in the next section.
 
\section{Results}
\label{sec:result}
For the numerical calculations we considered the masses of the scalars to be $m_h = 125$~GeV, $m_S$\, = 140 - 150~GeV, $m_H$\, = 250 - 270~GeV, and $m_A = m_{H^\pm}$\, = 400 - 600~GeV and $0.5\leq \tan\beta \leq 1.0$. In a few cases we have also scanned over light masses of $A$, and have discussed this where appropriate. Note also that the parameter space chosen here and in Ref.~\cite{Muhlleitner:2016mzt} are consistent with:
\begin{itemize}
\item[a)] theoretical constraints, such as vacuum stability from the global boundedness and minimum of the potential, as well as tree-level perturbative unitarity etc. 
\item[b)] experimental constraints from $B\to X_s \gamma$~\cite{Deschamps:2009rh,Mahmoudi:2009zx,Hermann:2012fc,Misiak:2015xwa} and from $R_b$~\cite{Haber:1999zh,Deschamps:2009rh}.
\item[c)] compatibility with $S$, $T$ and $U$, the oblique parameters.
\end{itemize}

As a first test, we start with 2HDMs where the positive one-loop contributions are from the terms involving $h$ and $H$, whereas the terms with $A$ and $H^{\pm}$ give negative contributions. Conversely, $h$ and $H$ provide the negative two-loop contributions, whereas $A$ provides a positive contribution. In the region of large $\tan\beta$ and small $m_A$, the 2HDM two-loop contributions become larger than the one-loop contributions, allowing for an explanation of the discrepancy. The required parameter space of the type-II and type-X 2HDMs is shown in~\cref{2HDM}. For a complete analysis of the 2HDM contributions see Ref.~\cite{Cherchiglia:2016eui}.

Although the addition of $S$ is necessary for the explanation of the multi-lepton anomalies, it does not have a significant effect on the parameter space of the 2HDM required to account for $\Delta a_\mu$, since the contributions from the CP-even scalars are suppressed with respect to that of the CP-odd scalar. This can be seen with a comparison of the plots in~\cref{2HDM} and the plots in~\cref{2HDM+S_high_tanbeta}. In contrast to the existing constraints on the 2HDM+S, ~\cref{2HDM+S_high_tanbeta} shows that it requires a high value of $\tan\beta$ and a relatively small mass of $A$. We have also shown the constraints on the parameter space in the $m_A$-$\tan\beta$ plane excluded by the direct searches at the LEP and the measurement of $\tau \to \mu \nu_\mu \nu_\tau$ at 95\% C.L. In all cases considered here, $\tau$-decays gives the most stringent constraints for both models and types. To calculate the leptonic $\tau$ decay we consider the leading one-loop diagrams and used the formula given in Ref.~{\cite{Abe:2015oca}}.

Following the model used in Ref.~\cite{vonBuddenbrock:2018xar}, where $m_A=600$ GeV and $\tan\beta < 1$, the full two-loop contribution from the 2HDM+S is shown in~\cref{2hdm+s_cont}. From the plots in~\cref{2hdm+s_cont} it is clear that the existing constraints on $\tan\beta$ and the particle masses in the 2HDM+S make the model unsuitable to account for the $\Delta a_\mu$. 

{\it Addition of BSM fermions}: Introducing additional fermionic degrees of freedom that interact with the 2HDM+S particles allows the one-loop contributions to become larger than the two-loop contributions. Fermions in the mass range 100~GeV to 1000~GeV enlarge the one-loop contributions enough to account for the $\Delta a_\mu$. The one-loop diagram with the new fermions is explained in~\cref{1&2-loop}. 
In~\cref{fermion_cont} we depicted the contribution to the $\Delta a_\mu$ by using~\cref{eqn:delta_a_f} for different choices of couplings. Since the contributions from the 2HDM+S without the BSM fermions and the two-loop diagrams containing the SM fermions are much smaller, only the one-loop contribution is considered here. In~\cref{fermion_couplings} we depict some of the couplings required by different fermion masses.

\section{Constraints and implications}
\label{constraint}
So far we constrained the parameter space of 2HDM+S and 2HDM+S+f models by calculating the contributions to $\Delta a_\mu$. However, there are other sources from where these model parameters can also be constrained. We discuss some of them in this section.
 
{\it Low energy bounds:} The observed data of $B_s \to \mu^+ \mu^-$ highly constrains the parameter space of 2HDMs and hence they also constrain the extended models considered here. Recall that the branching fraction of $B_s \to \ell \ell$ has an enhancement of $\tan^4 \beta$ in type~II 2HDMs, arising from the mediation of the extra Higgses now possible in the box and penguin type diagrams, as both charged leptons and down-type quarks have a coupling related to $\tan\beta$ for the extra bosons. However, the same couplings in a type~X 2HDM are enhanced by $\tan\beta$ for leptons and $\cot\beta$ for quarks, which implies no such $\tan^4 \beta$ relationship in the $B_s \to \ell \ell$ branching fraction in the type~X case. In fact, for type~X the leading contribution is largely $\tan\beta$ independent for large $\tan\beta$. A caveat to this (when comparing the type~X to the type~II 2HDM) is related to the contributions from $A$, the light CP-odd Higgs. $A$ can give large contributions to the branching fraction of $B_s \to \mu^+ \mu^-$, needed to explain the $(g-2)_\mu$ anomaly. From the $B_s \to \mu^+ \mu^-$ formula in Ref.~\cite{Logan:2000iv},
constraints on the parameter space are obtained in Ref.~\cite{Abe:2015oca}. A similar observation is allowed for the models and parameters considered in this work.

Furthermore, recent results from the CMS collaboration have presented the branching fraction as Br$(B_s \to \mu^+\mu^-) = [ 2.9^{+0.7}_{-0.6} ({\rm exp}) \pm 0.2 ({\rm frag})]\times10^{-9}$, where the first uncertainty combines the experimental statistical and systematic contributions, and the second is due to the uncertainty in the ratio of the $B_s^0$ and the $B^+$ fragmentation functions~\cite{Sirunyan:2019xdu}. This also conforms with our analysis.

{\it Muon mass corrections:} Due to the addition of singly charged SM singlet vector-like leptonic fermions given in~\cref{lagsf}, the mixing between muon and vector-like singlet leptonic fermion $f^{\prime}$ can cause corrections to the muon mass. Note that the quantum corrections from the vector-like leptons to the SM lepton masses are avoided by the minimal flavor protections~\cite{Freitas:2014pua}.
Here the Yukawa coupling between muon and the corresponding vector-like lepton $f^{\prime}$ will contribute to the muon mass proportional to the Yukawa coupling suppressed by the usual loop factor of ${\cal O}(1/(4\pi)$. Perturbativity and the accepted mass correction impose constraints on the relevant Yukawa couplings in the muon sector. In this work our choice of the couplings $|y_{f^\prime}^i|$ are limited by these requirements as shown in~\cref{fermion_cont}.

{\it Collider searches and limits:} The results provided by the ATLAS collaboration at center of mass energies of $\sqrt{s} = 13$~TeV have excluded singly charged ($L^\pm$) and neutral ($N^0$) heavy leptons with masses below 560~GeV at 95\% C.L., where the search was carried out in a simplified type-III seesaw model assuming branching fractions to all lepton flavors to be equal~\cite{ATLAS:2018ghc}. This translates to the limits on the $W^\pm L^\pm N^0$ coupling. 
Several type-III seesaw heavy lepton searches were also performed in the past by the ATLAS in Run~1 at $\sqrt{s} = 8$ TeV~\cite{Aad:2015cxa}, which excluded the heavy leptons with masses below 335~GeV. In Run~1 this search was complemented by another ATLAS search for heavy leptons using the three-lepton final state~\cite{Aad:2015dha}, excluding heavy lepton masses below 470~GeV. A Run~2 search by the CMS experiment at $\sqrt{s} = 13$~TeV~\cite{Sirunyan:2017qkz} was performed on multi-lepton final states using at least three leptons, excluding the type-III seesaw heavy leptons with masses in the range up to 840~GeV.

A similar analysis can be performed on the parameter space of the model considered here through a process $p p \to H \to f^\prime f^\prime$, where $f^\prime \to Z \mu^\pm$ or $f^\prime \to W^\pm \nu$ with final states as (semi-) leptonic and/or hadronic signatures with missing energy. In this work $m_S < 2\,m_{f^\prime}$, though similar search strategies can be followed with $A$. Here we can derive similar limits, where we should also note from earlier studies~\cite{Kumar:2016wzt, vonBuddenbrock:2016rmr} that resonance production of $S$ is suppressed and one should also consider $m_H \gtrsim 2\,m_{f^\prime}$, which limits $m_{f^\prime} \approx 150$~GeV (for on-shell decay) according to the parameter choices considered here. Following the limits derived from Ref.~\cite{ATLAS:2018ghc}, the couplings $H{f^\prime}{f^\prime}$ and/or $A{f^\prime}{f^\prime}$ should be in the range of $\sim [0.4,1.0]$ in order to observe such final states. These analyses further strengthen the parameter space to explain $\Delta a_\mu$.
Here we discussed a search strategy roughly which can be taken as future work for full analyses.       

\section{Summary and conclusion}
\label{sec:concl}

A number of predictions were made in Refs.~\cite{vonBuddenbrock:2015ema,vonBuddenbrock:2016rmr} pertaining to the anomalous production of multiple leptons at high energy proton-proton collisions. These could be connected with a heavy boson with a mass around the electroweak scale decaying predominantly into a SM Higgs boson and a singlet scalar. Discrepancies in multi-lepton final states were reported with Run~1 data in Refs.~\cite{vonBuddenbrock:2015ema,vonBuddenbrock:2017gvy} and have now become statistically compelling with the available Run~2 data~\cite{vonBuddenbrock:2019ajh}. These include the production of opposite-sign, same-sign and three leptons with and without $b$-quarks. Discrepancies arising in final states and parts of the phase-space where different SM processes can dominate, points to the unlikeliness of a mis-modeling of one particular SM process, and has led to the anomalies and their kinematic characteristics being well described by a simple ansatz. This ansatz is that a $H$ with mass of $m_H\approx 270$\,GeV, produced in association with top quarks via gluon fusion, can decay via $H\to Sh$, where $m_S\approx 150$\,GeV. The $H$ can be embedded into a 2HDM, whilst the $S$ can be an additional singlet (a 2HDM+S)~\cite{vonBuddenbrock:2016rmr,Muhlleitner:2016mzt,vonBuddenbrock:2018xar}. The 2HDM+S model which accommodates all these features 
of the data~\cite{vonBuddenbrock:2018xar} are used here as a baseline.

The long-standing discrepancy in the muon anomalous magnetic moment are explored here in connection with the suggested scalar boson spectroscopy and in the context of a constrained 2HDM+S model. The two-loop contribution from the 2HDM+S has been calculated, and it has been shown that this contribution is too small to account for the $\Delta a_\mu$  discrepancy. In addition, the values of $\tan\beta$ and $m_A$ required to explain the discrepancy within 2 sigma have been determined for the 2HDM+S. In both the type- II and type-X 2HDM+S, a light pseudo-scalar is required, along with a high value of $\tan\beta$. This choice of parameters in the 2HDM+S model is not compliant with the features of the LHC data described here. In order to be able to explain the $\Delta a_\mu$ discrepancy with the 2HDM+S model as constrained in Ref.~\cite{vonBuddenbrock:2018xar}, additional BSM fermionic degrees of freedom may be required. Given the size and the errors associated with the $\Delta a_\mu$  anomaly, new leptons would need to be as heavy as $\mathcal{O}(100)$\,GeV. The impact of these new degrees of freedom on the model considered here and, in particular, on the decays of bosons is beyond the scope of this paper and will be covered in subsequent works, though we also discussed the search strategies in brief with other constraints in section~\ref{constraint}.

\section{acknowledgements}
The authors are grateful for the support from the South African Department of Science and Innovation through the SA-CERN program and the National Research Foundation for various forms of support.

\end{document}